\documentclass[rnote]{aa}
\usepackage{graphicx}
\usepackage{natbib}
\bibpunct{(}{)}{;}{a}{}{,}
\usepackage{longtable}
\begin{document} 

\title{The Hvar survey for roAp stars: II.  Final results\thanks{Figures 1 and 2 
are only available in electronic form via http://www.edpsciences.org}}
\author{E.~Paunzen\inst{1} 
\and M.~Netopil\inst{1}
\and M.~Rode-Paunzen\inst{2}
\and G.~Handler\inst{3}
\and H.~Bo\v{z}i\'{c}\inst{4}}


\institute{Department of Theoretical Physics and Astrophysics, Masaryk University,
Kotl\'a\v{r}sk\'a 2, 611\,37 Brno, Czech Republic
\email{epaunzen@physics.muni.cz}
\and Institut f{\"u}r Astrophysik der Universit{\"a}t Wien, T{\"u}rkenschanzstr. 17, A-1180 Wien, Austria
\and Copernicus Astronomical Center, Bartycka 18, 00-716, Warsaw, Poland
\and Hvar Observatory, Faculty of Geodesy, University of Zagreb, Ka\v{c}i\'{c}eva 26, HR-10000 Zagreb, Croatia}

\date{}

\abstract
{The 60 known rapidly oscillating Ap (roAp) stars are excellent laboratories to test pulsation models
in the presence of stellar magnetic fields. Our survey is dedicated to search for new group members in the 
Northern Hemisphere.}
{We attempt to increase the number of known chemically peculiar stars that are known to be pulsationally unstable.}
{About 40\,h of new CCD photometric data of 21 roAp candidates, observed at the 1\,m Austrian-Croatian Telescope (Hvar
Observatory) are presented. We carefully analysed these to search for pulsations in the frequency range of up to 10\,mHz.}
{No new roAp star was detected among the observed targets. The distribution of the upper limits for roAp-like variations
is similar to that of previoius similar efforts using photomultipliers and comparable telescope sizes.}
{In addition to photometric observations, we need to consolidate spectroscopic information to select suitable targets.}
\keywords{Stars: chemically peculiar -- variables: general}
\maketitle

\section{Introduction}

The rapidly oscillating Ap (roAp) stars can be 
found within an area of the pulsational instability
in the Hertzsprung-Russell diagram,
between the zero-age main-sequence and terminal-age main-sequence, 
ranging in effective temperature from about 6600\,K to 8500\,K.  
They are characterized by overabundances of 
up to several dex of, for example, 
strontium, chromium, europium, and other rare-earth elements when compared to the Sun.
The roAp stars show pulsational periods in the range of about 5 to
25 minutes with amplitudes of up to 10\,mmag in Johnson $B$.

Since their first detection by \citet{Kurt82}, about 60 stars of this type have 
been discovered. Almost all members of this group have been found with ground-based photometric 
observations of known classical chemically peculiar (CP) stars. With the new high-precision 
space-based data, several new low-amplitude roAp stars have been detected \citep[][and references therein]{Hold14}. 
In addition to this, some roAp stars have been discovered by spectral line variations of elements 
such as cerium, neodymium, and samarium \citep{Elki11}.

Most, but not all, of the roAp stars exhibit strong global magnetic fields with
values of up to 25\,kG \citep{Hubr12}. Therefore, these stars are excellent test cases
for studying the interactions between magnetic fields and stellar pulsation. 
The driving mechanism of their
oscillations is most probably the classical
$\kappa$-mechanism operating in the hydrogen ionisation zone \citep{Balm01}. 
Recently, \citet{Cunh13} suggested another excitation mechanism,
however, where the pulsations 
are driven by the turbulent pressure in the convection zone.

We initiated a photometric survey to search for northern roAp stars at the Hvar observatory
\citep[][ Paper I]{Paun12}. Together with the new observations, we collected
100 hours of time series in Bessell $B$ for 41 individual targets. Observations of three targets (Renson 1860, 58275, and
58777) were presented in both papers, but Renson 59590 is misclassified as a CP star. This left us with 40 potential
roAp candidates. The accuracy of our data
is similar to those of former, similar, efforts \citep{Josh06}. We did not detect any new roAp star, but were able to establish upper limits for variability.

\begin{table*}
\begin{center}
\caption[]{Basic data of the new target stars (upper panel), previously observed stars from Paper I (lower panel), and the results of the time-series
analysis.}
\begin{tabular}{lccccccccc}
\hline
\hline
Renson & HD/BD/Tycho & $V$   & Spec\tablefootmark{a} & $T_\mathrm{eff}$ & $\sigma T_\mathrm{eff}$\tablefootmark{b}& $E(B-V)$ & JD(start)\tablefootmark{c} & $\Delta t$ & UL\tablefootmark{d} \\
       &             & [mag] &                       & [K]              & [K]                     & [mag]    & [d]          & [min]      & [mmag] \\
\hline        
680 & 2852 & 8.954 & A5 Sr Eu & 8700 & 190 (3) & 0.16 & 1888.57737 & 87.4 & 2.0 \\
45762 & 162162 & 9.328 & A3 Sr Eu & 7410 & 250 (4) & 0.15 & 1892.27993 & 97.9 & 2.3 \\
45763 & 162177 & 8.645 & A4 Sr Eu & 8190 & 310 (4) & 0.22 & 1888.27779 & 60.9 & 2.5 \\
47940 & +17 3622 & 8.821 & A2 Sr Eu Cr & 8720 & 160 (3) & 0.01 & 1895.27950 & 96.6 & 2.5 \\
48750 & 174021 & 8.977 & F -- Sr & 7960 & 180 (3) & 0.15 & 1887.29557 & 90.5 & 2.0 \\
49170 & 337282 & 9.419 & A0 Si & 9440 & 300 (3) & 0.10 & 1885.33120 & 130.5 & 1.7 \\
49173 & 176281 & 8.230 & F2 Sr & 6860 & 400 (3) & 0.02 & 1896.27083 & 90.6 & 2.0 \\
49830 & +44 3074 & 9.944 & A3 Si & 8820 & 380 (4) & 0.10 & 794.31381 & 70.0 & 4.0 \\
        &               &               &               &               &               &               & 802.29204 & 131.1 & 1.3 \\
        &               &               &               &               &               &               & 835.32726 & 64.7 & 3.2 \\
        &               &               &               &               &               &               & 837.23579 & 38.8 & 2.3 \\
        &               &               &               &               &               &               & 838.23175 & 73.1 & 2.2 \\
        &               &               &               &               &               &               & 839.29245 & 66.2 & 2.1 \\
        &               &               &               &               &               &               & 841.29166 & 68.2 & 1.8 \\
        &               &               &               &               &               &               & 1888.35615 & 93.7 & 2.4 \\
50160 & 344100 & 9.909 & A3 Si & 8540 & 220 (3) & 0.01 & 1887.36769 & 96.9 & 2.0 \\
50600 & +35 3616 & 9.477 & F0 Sr Eu & 6830 & 410 (3) & 0.08 & 799.33105 & 134.0 & 3.2 \\
50924 & 234924 & 9.473 & A2 Sr & 9460 & 140 (3) & 0.00 & 1892.35595 & 95.3 & 2.3 \\
52200 & 339199 & 10.142 & A0 Si Sr & 9410 & 660 (3) & 0.27 & 1888.42960 & 92.7 & 2.6 \\
52670 & 189919 & 8.980 & A Si & 10180 & 490 (3) & 0.00 & 1887.44274 & 86.0 & 1.9 \\
53542 & 192060 & 8.762 & A5 Y Sr & 7770 & 230 (4) & 0.14 & 1895.36030 & 94.7 & 2.2 \\
54440 & 332312 & 9.722 & A4 Sr & 8200 & 580 (3) & 0.17 & 1886.47754 & 144.4 & 1.6 \\
54800 & 196542 & 8.986 & A4 Sr Cr Eu & 8620 & 340 (4) & 0.05 & 1895.43558 & 96.3 & 2.4 \\
55130 & 341037 & 9.435 & F0 Sr Cr Eu & 8040 & 160 (3) & 0.03 & 799.39451 & 130.6 & 2.1 \\
59590\tablefootmark{e} & +38 4163 & 11.123 & A7 Sr Si & & & & 1887.51775 & 67.8 & 3.1 \\
59910 & +46 3884 & 9.137 & F0 Cr Eu & 7660 & 220 (4) & 0.14 & 1888.50305 & 93.9 & 2.4 \\
59980 & +62 2151 & 9.805 & A1 Sr & 9370 & 250 (3) & 0.21 & 1895.51014 & 73.3 & 1.7 \\
60972 & +51 3678 & 9.984 & A2-- & 8910 & 210 (3) & 0.00 & 1887.57091 & 90.7 & 2.2 \\
\hline
1860    &       7410    &       9.076   &       A5 Sr Cr Eu     &       7920 &       160 (3) &       0.02    &       837.49306       &       134.6   &       1.9     \\
        &               &               &               &               &               &               &       838.48095       &       70.1    &       2.7     \\
        &               &               &               &               &               &               &       839.43888       &       136.8   &       1.9     \\
        &               &               &               &               &               &               &       841.45374       &       72.0    &       2.6     \\
58275   &       +46 3543        &       9.728   &       A2 Si   &       8660    &       340 (3)     &       0.08    &       837.39956       &       94.7    &       3.2     \\
        &               &               &               &               &               &               &       838.41903       &       69.6    &       2.5     \\
        &               &               &               &               &               &               &       839.35095       &       106.6   &       2.4     \\
        &               &               &               &               &               &               &       841.34966       &       70.7    &       3.2     \\
58777   &       3982-4172-1     &       10.737  &       A3 Sr   &       7930    &       410 (3)     &       0.19    &       839.35921       &       73.9    &       2.9     \\
\hline
\end{tabular}
\tablefoot{
\tablefoottext{a}{\citet{Rens09}} 
\tablefoottext{b}{The number of temperature calibrations used are given in parenthesis.}
\tablefoottext{c}{JD$-$2\,455\,000} 
\tablefoottext{d}{Upper limit in Figs. \ref{fourier_1} and \ref{fourier_2}} 
\tablefoottext{e}{. Probably misclassified, see text.}
}
\label{stars}
\end{center}
\end{table*}

\section{Target selection, observations, and reduction}

All targets were selected from the catalogue of Ap, HgMn, and Am stars
by \cite{Rens09}. Although most known roAp stars are classified as
SrCrEu, we widened our list of 
targets to also include hotter spectral types, that is, silicon stars, to avoid missing
any variables due to a bias in the
selection process. In total, we selected 21 further objects for observations.

All observations were performed
at the Hvar Observatory, University of Zagreb, using the 1\,m
Austrian-Croatian Telescope (ACT) with the following equipment:

\begin{itemize}
\item August, September, and October 2011: Apogee Alta U47 CCD camera, 1024x1024 pixels, 
a field of view of about 3$\arcmin$,
\item August 2014: Moravian instrument, G2-1600 KAF1603ME CCD camera, 1536x1024 pixels,
a field of view, using a focal reducer, of about 10$\arcmin$x8$\farcm$
\end{itemize}

The integration times for the observations in the Bessell $B$ filter system
were set to between 10 and 45 seconds, mainly depending on the brightness of the target
and comparison stars as well as the seeing. 

The data reduction and differential photometry were performed using the C-Munipack
package\footnote{http://c-munipack.sourceforge.net/}. If several comparison stars 
were available, these were checked individually to exclude variable objects.
We compared the results of the
final differential light curves using the aperture photometry routine from 
IRAF\footnote{Available from http://iraf.noao.edu/}. We found no differences above the
photon noise.

The final light curves were examined in more detail 
using the program Period04 \citep{Lenz05}, which performs a discrete Fourier transformation. 
Significant peaks exceeding the noise level of more than 4$\sigma$
with periods of more than one hour were subtracted. The results were checked
with those from the phase-dispersion method computed 
within the program package Peranso\footnote{http://www.peranso.com/}. No significant differences
were noticed.

The detailed observational dates and results of the
Fourier time-series analysis for all targets are listed in 
Table \ref{stars} and shown in Figs. \ref{fourier_1} and \ref{fourier_2}.

\onlfig{
\begin{figure*}
\begin{center}
\includegraphics[width=160mm]{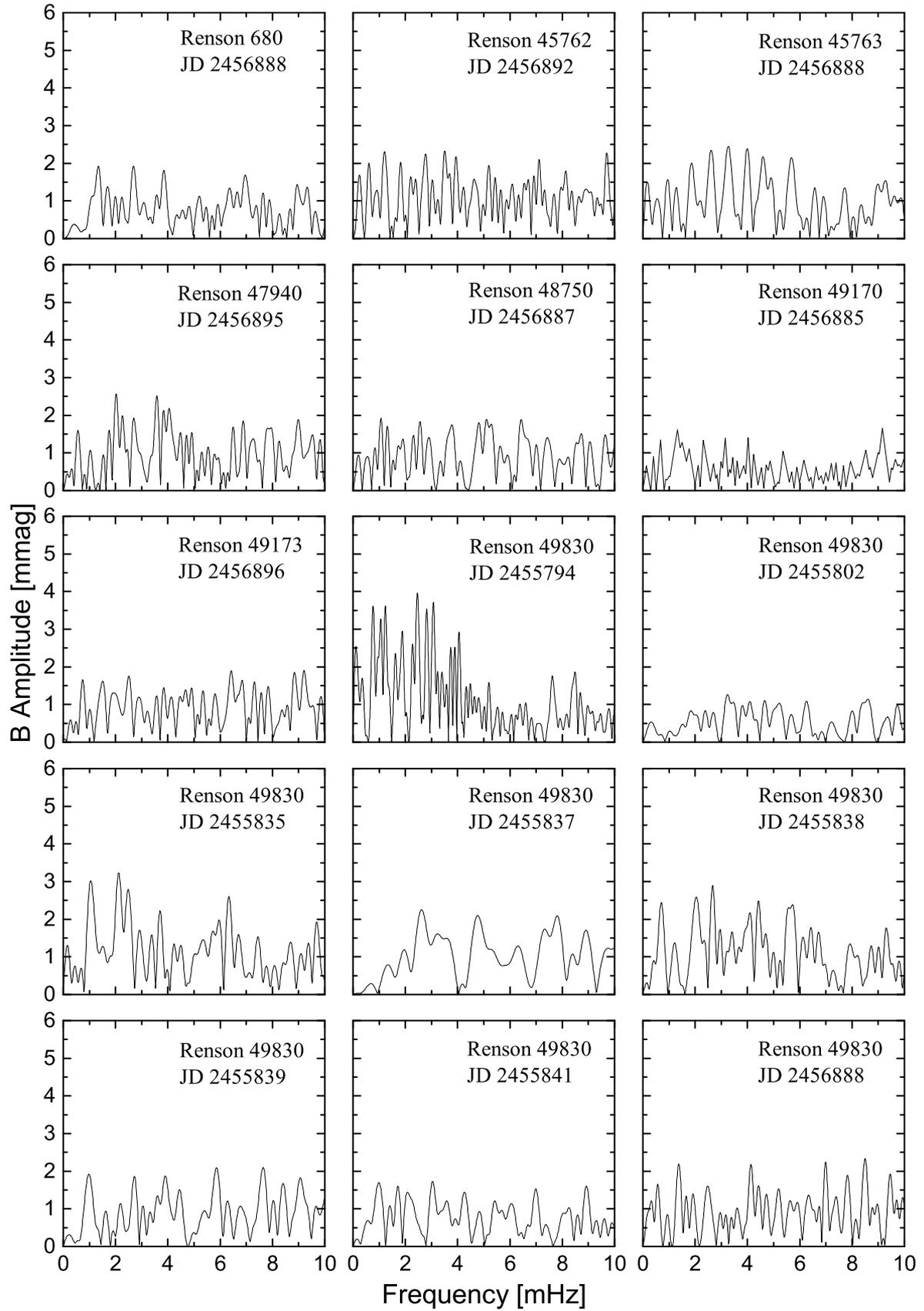}
\caption[]{Fourier spectra of target star light curves first investigated for roAp oscillations in this paper.}
\label{fourier_1}
\end{center}
\end{figure*}
\addtocounter{figure}{-1}
\begin{figure*}
\begin{center}
\includegraphics[width=160mm]{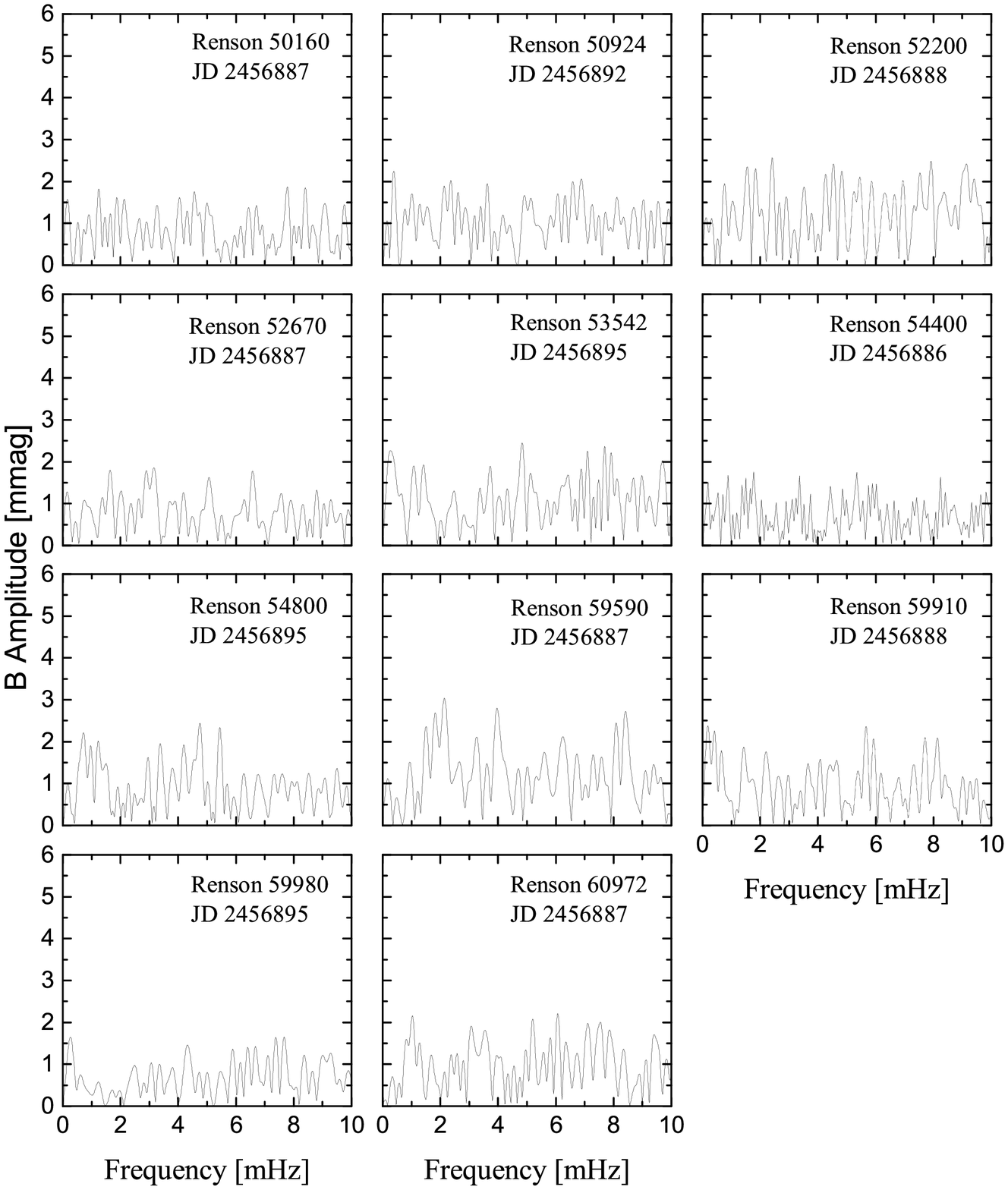}
\caption[]{continued.} 
\end{center}
\end{figure*}

\begin{figure*}
\begin{center}
\includegraphics[width=160mm]{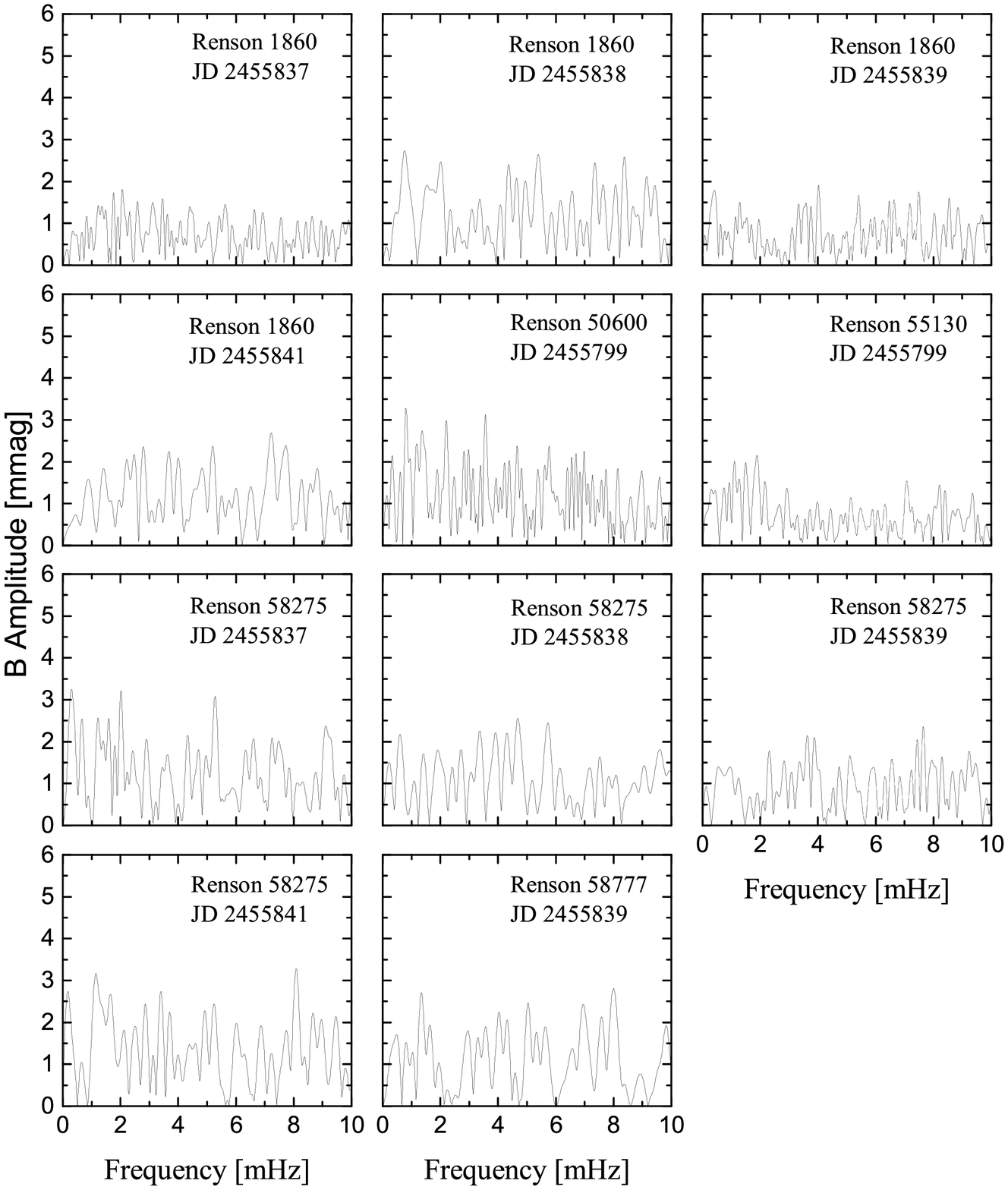}
\caption[]{Fourier spectra of the target star light curves previously investigated for roAp oscillations.} 
\label{fourier_2}
\end{center}
\end{figure*}
}

\addtocounter{figure}{+1}
\begin{figure}
\begin{center}
\includegraphics[width=80mm]{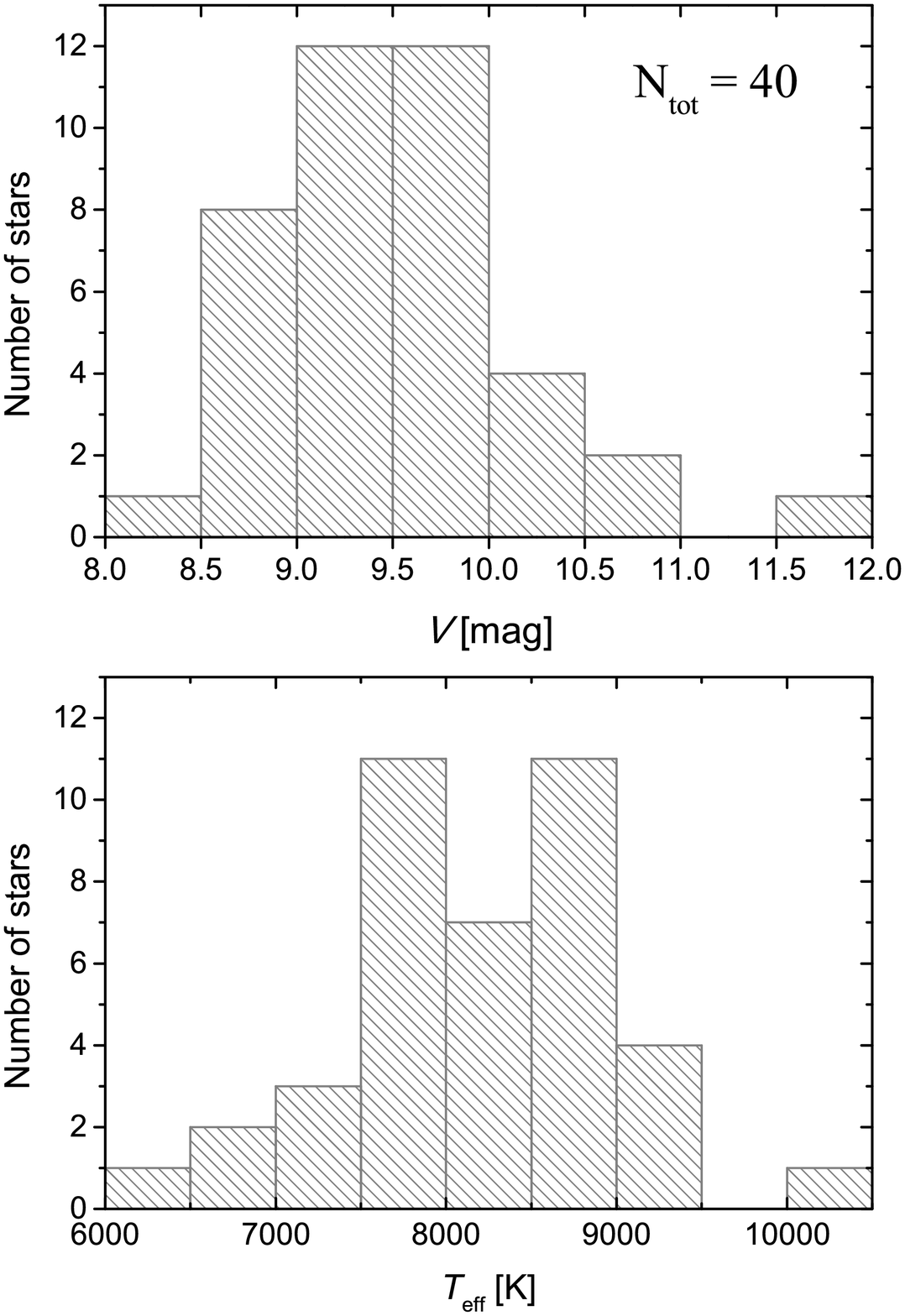}
\caption[]{Distributions of the brightness and effective temperature for the 40 investigated targets.}
\label{stat_1}
\end{center}
\end{figure}

\begin{figure}
\begin{center}
\includegraphics[width=80mm]{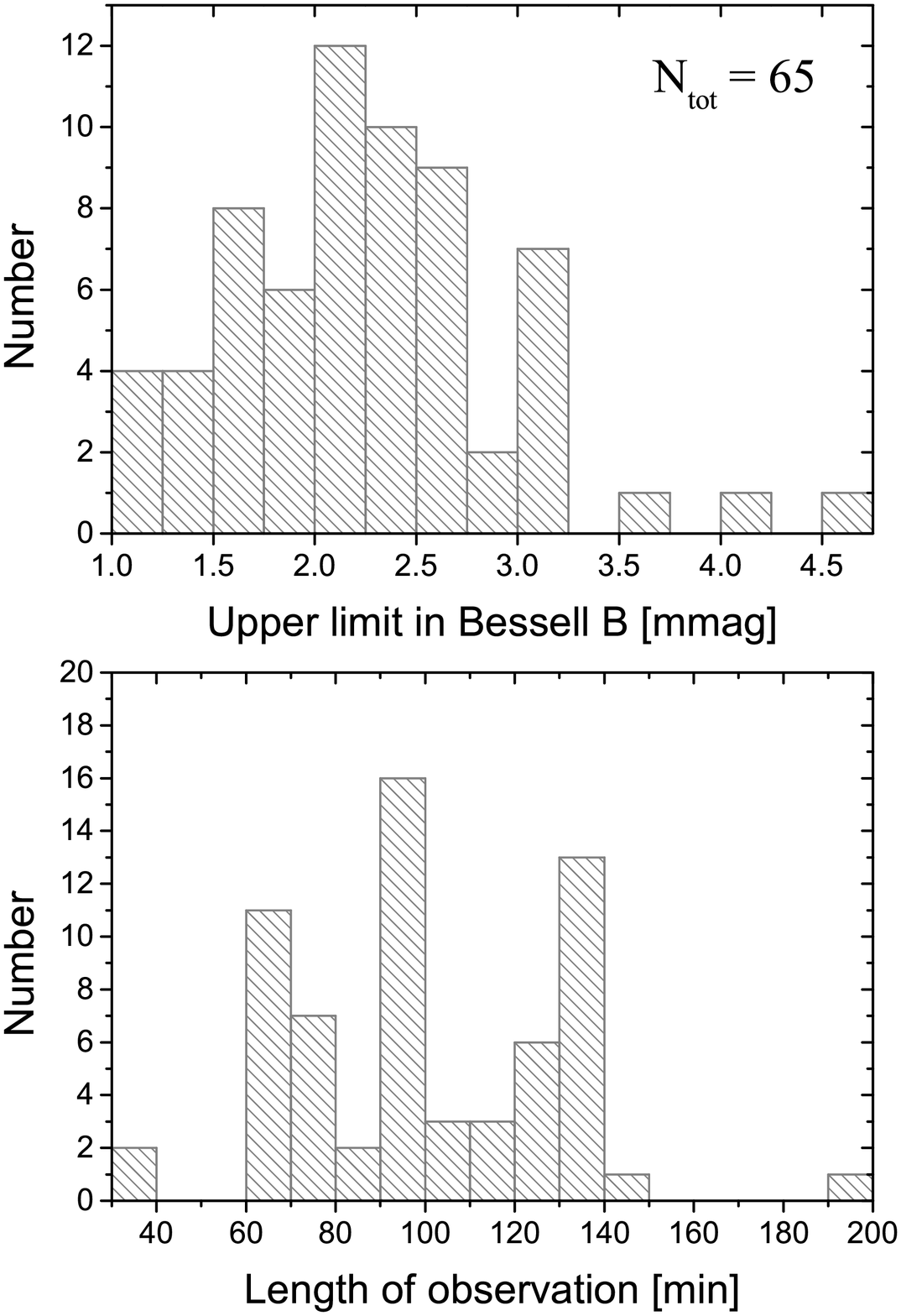}
\caption[]{Distributions of the upper limits for roAp-like variations and the length of observation for
the 65 individual data sets.}
\label{stat_2}
\end{center}
\end{figure}

\begin{figure}
\begin{center}
\includegraphics[width=80mm]{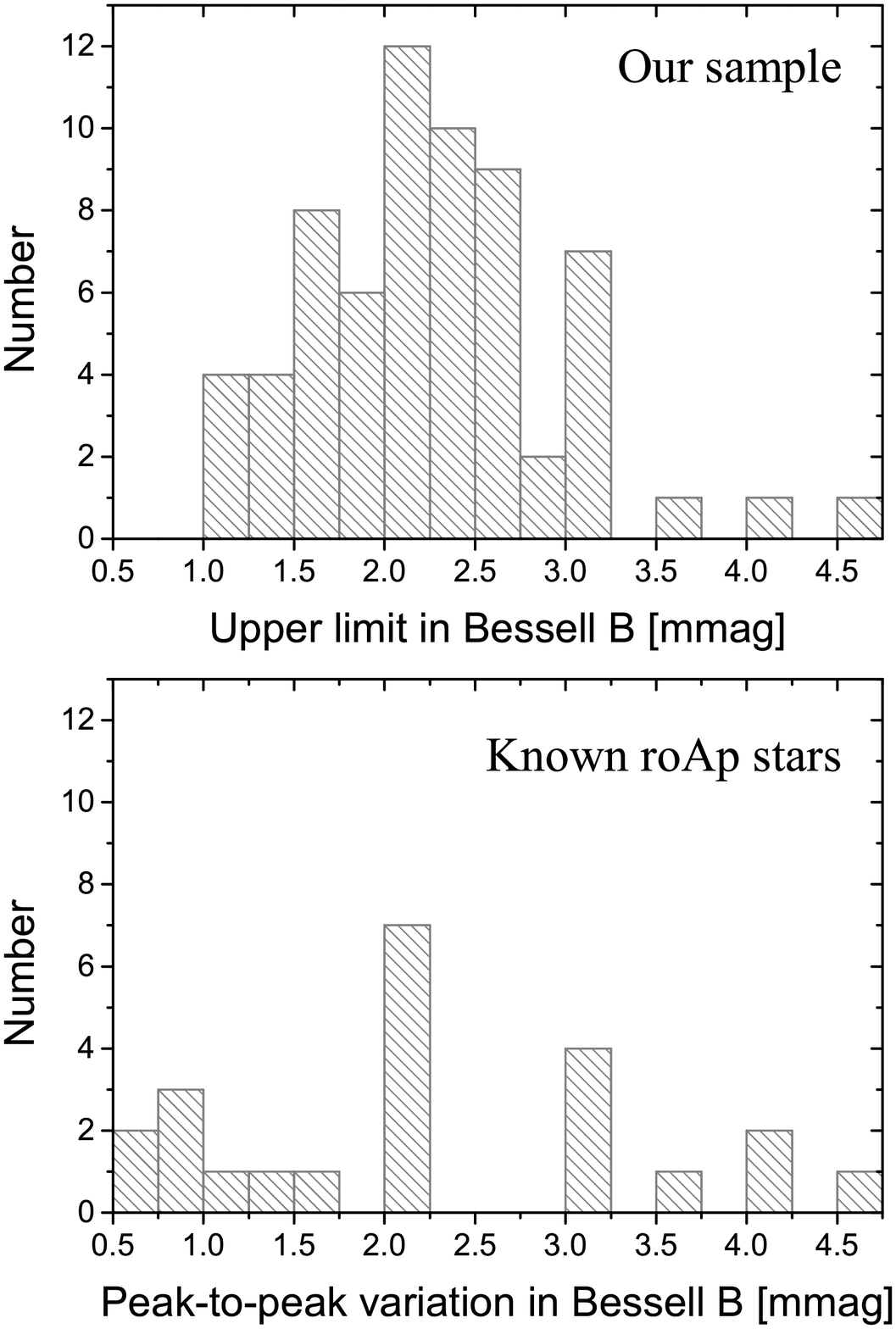}
\caption[]{Distributions of the upper limits for roAp-like variations from our sample and the peak-to-peak
variation of known roAp stars from \citet{Kurt06}.}
\label{stat_3}
\end{center}
\end{figure}

\section{Analysis and discussion}

Figure \ref{stat_1} shows the distributions of the apparent brightness and effective temperature for the 40 investigated targets.
The effective temperatures were calibrated as described in Paper I. In short, they are mean values from different photometric
calibrations and fitting the individual spectral energy distributions (SED). Almost all targets are between 8.5 and 10.5 magnitude 
in Johnson $V$. Taking into account the reddening values from Table \ref{stars} and corresponding isochrones \citep{Bres12},
a mean distance modulus of 8\,mag (or a distance of 400\,pc) places all targets between the zero- and terminal-age main-sequence.
The range of effective temperatures is broader than the values of known roAp stars.

The distributions of the upper limits for roAp-like variations and the length of observation for
the 65 individual data sets (about 100 hours, in total) are shown in Fig. \ref{stat_2}. We can directly compare them with those
of the null-results from  the Nainital-Cape survey \citet[][Fig. 7 therein]{Josh06}. They used
a Johnson $B$ filter, a photoelectric detector, and a 1.04\,m telescope for their photometric time-series observations. 
The upper limits range from 0.3 to 5.3\,mmag with a peak at about 1.3\,mmag. The peak of our distribution is
shifted by about 1\,mmag to higher values. 
This is probably caused by the lower quantum efficiencies of the 
CCDs compared to the photomultiplier at the same integration time (limited by the sufficient coverage of the 
highest known roAp pulsation frequency) and telescope size. In addition, the observations were performed at a much
higher altitude (about 2000\,m, while Hvar observatory is at 238\,m above the sea level), which results in a lower scintillation noise.
The length of observations guarantee that at least three full pulsation cycles are covered, taking into account the longest known pulsation period of 23.5\,min of the roAp star
HD 60435 \citep{Kurt06}. 

We compared the results of the time series limits with the distribution of the peak-to-peak variations of known roAp
stars from \citet{Kurt06}. The latter are also for $B$ magnitudes. 
Additional later detected roAp stars were either found spectroscopically \citep[e.g.,][]{Koch13} or are low-amplitude stars in a different filter
\citep[e.g.,][]{Hold14}.
No attempt was made to transfer
amplitudes from other photometric system to $B$ magnitudes. Figure \ref{stat_3} shows the comparison of both distributions. Nine known
roAp stars exceed a peak-to-peak variation of 5\,mmag. Therefore, we
conclude that there might be bona-fide roAp stars among our sample that were not detected because of their low photometric amplitudes
(several members do not even show photometric variation at all).

In the following, we discuss the results for some individual stars in more detail.

{\it Renson 1860} and {\it 58275:} results for both stars were presented in Paper I. We re-observed them
on four nights, each, to investigate their long-term (in)stability. \citet{Balo13} showed that
the amplitudes of the individual frequencies can vary significantly over time also with the
rotational phase, for example. In our case, this would mean
that we miss possible statistically significant amplitudes. However, we were unable to find a positive 
detection.

{\it Renson 45762} and {\it 45763:} both stars are located in the area of the open cluster IC 4665 (age about 
40\,Myr). According to \citet{Zejd12}, the mean proper motion of IC 4665 is, based on five individual references, 
$\mu_{\alpha}\cos \delta$\,=\,$-$0.8$\pm$0.2\,mas\,yr$^{-1}$
and $\mu_{\delta}$\,=\,$-$8.3$\pm$0.1\,mas\,yr$^{-1}$, respectively. The proper motions of the targets are
[+22.8$\pm$1.8,$-$4.6$\pm$1.7] and [+1.9$\pm$1.9,+8.9$\pm$1.9] taken from the Tycho-2 catalogue \citep{Hog00}.
We conclude that neither star is a member of IC 4665.

{\it Renson 49170:} \citet{Rens09} gave an incorrect identification (BD+27\,3185), the correct one is
BD+27\,3184.

{\it Renson 49830:} for this object, we analysed eight different data sets because the first one showed a
very high noise level (about 4\,mmag) in the interesting frequency domain. In the consecutive observations,
the upper limits range from 1.3 to 3.2\,mmag. In none of these observations did we find a statistically significant
peak. 

{\it Renson 52200:} in Simbad/CDS the reference for `Renson 52200' is incorrect. 
It is linked to the star HD 339199, but with incorrect coordinates, as also given by \citet{Rens09}. The correct star, which appears 
brighter and bluer, is about one arcminute away and currently listed as HD 399199 in Simbad. 
This ID is most likely a typographical error in Simbad/CDS, because the number
exceeds the entries in the HD catalogues. 
The star is actually HD 339199 according to the catalogue by \citet{Fab02}, for example.

{\it Renson 54800:} This is a very interesting close binary system (separation of about 3$\arcsec$) consisting of
a CP1 and a CP2 star \citep{Abt84}, for which \citet{Cowl64} reported possible spectrum variability. 
The two components were not resolved, therefore our observations are for the combined flux.

{\it Renson 58777:} In Paper I, we stated that this is a good candidate for follow-up observations. Here, we
present one additional data set (Fig. \ref{fourier_2}). No statistically significant variability with an
upper limit of 2.9\,mmag was detected. However, this limit is just 0.1\,mmag lower than previously.
Therefore, this star is still worthy of further follow-up observations.

{\it Renson 59590:} From the available photometry we conclude that this object is an early G-type dwarf.
But it was classified as `A7 Sr' by \citet{Ziri51}. He listed a photographic magnitude
of 8.5\,mag, while $B/V$ is 12.09/11.12\,mag, respectively. We checked some other listed magnitudes and 
compared them to more recent magnitudes for stars with similar spectral types 
(for example, HD 188854) and found an excellent agreement. In the vicinity of Renson 59590, there are no other bright stars of eigth to
ninth magnitude, however. We conclude that there is probably a typographical error of the stellar designation in \citet{Ziri51}. This 
object is certainly no CP star.

{\it Renson 59980:} The star is located in the area of the Cep OB3 association, but very probably a nonmember based on kinematic and 
photometric data \citep{Kharch04}. The reddening of the star is lower than that of other members of the association \citep[e.g.,][]{Garrison70}, which confirms this classification.

\section{Conclusions}

We presented our efforts to search new northern roAp stars in two papers. In total, 65 individual data sets with 
about 100 hours of CCD photometry for 40 stars were analysed without detecting a statistically significant
signal. The upper limits for the variability range are between 1 to 5\,mmag with a peak at about 2\,mmag.

Further high-precision photometric observations are needed to detect new roAp stars to initiate spectroscopic
follow-up observations to model their stellar atmospheres in more detail \citep{Nesv13}. As a first step, 
however, the only source for good candidate targets, the catalogue by \citet{Rens09}, has to be consolidated and
updated. In addition, even for bright stars, we lack modern classification resolution spectroscopy
\citep{Paun11}. We will continue our efforts to contribute to these different tasks in the future.

\begin{acknowledgements}
This project is financed by the SoMoPro II programme (3SGA5916). The research leading
to these results has acquired a financial grant from the People Programme
(Marie Curie action) of the Seventh Framework Programme of EU according to the REA Grant
Agreement No. 291782. The research is further co-financed by the South-Moravian Region. 
It was also supported by the grants GP14-26115P, 7AMB14AT015,
the financial contributions of the Austrian Agency for International 
Cooperation in Education and Research (BG-03/2013 and CZ-09/2014). 
GH is grateful for support by the Polish NCN grant 2011/01/B/ST9/05448.
HB acknowledges financial support by Croatian Science Foundation under
the project 6212 ``Solar and Stellar Variability''.
This research has made use of the WEBDA database, operated at the 
Department of Theoretical Physics and Astrophysics of the Masaryk University.
This work reflects only the authors' views, and the European 
Union is not liable for any use that may be made of the information contained therein.
\end{acknowledgements}

\end{document}